\documentclass[aps,prb,twocolumn,showpacs,superscriptaddress]{revtex4}

\usepackage{mathrsfs}
\usepackage{epsfig}
\usepackage{graphicx}
\usepackage{amsfonts}
\usepackage[figuresright]{rotating}
\usepackage{amssymb}
\usepackage{amsmath}
\usepackage{dcolumn}
\usepackage{bm}

\def\be{\begin{equation}}
\def\ee{\end{equation}}
\def\bea{\begin{eqnarray}}
\def\eea{\end{eqnarray}}

\def\nn{\nonumber}

\begin{document}
\title{
Magnetic phases of bosons with synthetic spin-orbit coupling
in optical lattices}
\author{Zi Cai}
\affiliation{Department of Physics, University of California, San
Diego, California 92093}
\author{ Xiangfa Zhou}
\affiliation{
Key Laboratory of Quantum Information, University of Science and Technology
of China, CAS, Hefei, Anhui 230026, China
}
\author{Congjun Wu}
\affiliation{Department of Physics, University of California, San
Diego, California 92093}
\affiliation{
Center for Quantum Information, IIIS, Tsinghua University, Beijing, China}

\begin{abstract}
We investigate magnetic properties in the superfluid and
Mott-insulating states of two-component bosons with spin-orbit (SO)
coupling in 2D square optical lattices. The spin-independent hopping
integral $t$ and SO coupled one $\lambda $are fitted from band
structure calculations in the continuum, which exhibit oscillations
as increasing SO coupling strength. The magnetic superexchange model
is derived in the Mott-insulating state with one-particle per-site,
characterized by the Dzyaloshinsky-Moriya (DM) interaction. In the
limit of $|\lambda|\ll |t|$, we find a spin spiral Mott state whose
pitch value is the same as that in the incommensurate superfluid
state, while in the opposite limit $|t| \ll |\lambda|$, the ground
state can be found by a dual transformation to the  $|\lambda|\ll
|t|$ limit.
\end{abstract}
\maketitle

Quantum many-body states with spontaneous incommensurate modulated structure
have attracted considerable interests in the past decades, and occur in many
settings of condensed matter and ultracold atom physics, such as frustrated
magnetism, unconventional superconductor and superfluid and so on.
Celebrated examples include the incommensurate magnetism with long-range
and short range magnetic order \cite{read1991,sachdev1991}, the
Fulde-Ferrell-Larkin-Ovchinnikov (FFLO) pairing states
\cite{fulde1964,larkin1965}.
Recently, the Bose-Einstein condensations (BEC) with spin-orbit (SO)
coupling introduce a new member to this family.
The SO coupled BECs are genuinely new phenomena due to the fact that
the kinetic energy is not just a Laplacian but also linearly depends
on momentum, which gives rise the complex-valued condensate
wavefunctions beyond Feynman's no-node theorem \cite{feynman1972}.

An interesting property of SO coupled condensates of bosons is that
they can spontaneously break time-reversal symmetry which is absent
in conventional BECs of both superfluid $^4$He and many experiments
of ultra-cold alkali bosons \cite{leggett2001}.
For example, it is predicted that such condensates can
spontaneously develop half-quantum vortex coexisting with 2D skyrmion-type
spin textures in the harmonic trap \cite{wu2008}.
Experimentally, spin textures of the SO coupled bosons have been observed in
exciton condensations, which is a solid state boson systems with
relativistic SO coupling \cite{high2012}.
Theoretically, extensive studies have been performed for SO coupled
bosons which exhibit various spin orderings and textures from
competitions among SO coupling, interaction, and confining trap
energy{\cite{wu2008,stanescu2008,ho2011,yip2011,wang2010,zhang2012,
zhou2011,hu2012,li2012a}.

In the optical lattice, the SO coupled bosons are even more
interesting. Early investigations have showed that the
characteristic incommensurate wavevectors are incommensurate with
the lattice \cite{mondragon2010}. In this article, we study the SO
coupled Bose-Hubbard model, focusing on the magnetic properties. The
tight binding model is constructed and the spin-independent hopping
integral $t$ and SO coupled hopping integral $\lambda$ are
calculated as functions of the SO coupling strength in the
continuum. Magnetic superexchange models are derived characterized
by the Dzyaloshinsky-Moriya (DM) interaction
\cite{dzyaloshinsky1958,moriya1960}. In the Mott-insulating phase,
single particle condensation is suppressed but the spin order is
not. The spin orderings are solved in two different limits,
$|\lambda|\ll |t|$, and $|t|\ll |\lambda|$, respectively. In the
former case, the DM term destabilizes the ferromagnetic state to
spin spirals, while in latter case can be transformed to the former
one by a dual transformation .

We begin with the non-interacting Hamiltonian of bosons with the Rashba SO
coupling in a square lattice optical potential as
\bea
H_0&=&\frac{\hbar^2\mathbf{k}^2}{2m}\hat{1} +
\frac{\hbar^2k_{so}}{m}(\alpha k_x\hat{\sigma}_y
+\beta k_y\hat{\sigma}_x)+V(x,y), \ \ \,
\label{eq:Ham0}
\eea
where $k_{so}$ is the magnitude of wavevectors of laser beams generating
SO coupling.
$\alpha$ and $\beta$ characterize the anisotropy of SO coupling.
Below we consider two situations.
First, SO coupling is only along the $x$-direction, i.e.,
$\alpha=1$, $\beta=0$, which agrees with the recent experiments\cite{lin2011}.
Second, the isotropic Rashba SO coupling with $\alpha=1$, $\beta=1$.
$V(x,y)$ is the periodic potential produced by laser beams with wavelength
$\lambda_0$ as
\bea
V(x,y)=-V_0[\cos^2(k_0x)+\cos^2(k_0y)]
\label{eq:op}
\eea
where $k_0=2\pi/\lambda_0$, and the recoil energy $E_r=\frac{\hbar^2k_0^2}{2m}$.
We define a dimensionless parameter
$\gamma_0=k_{so}/k_0$ to characterize the strength of SO coupling.
The lattice constant $a=\lambda_0/2$, and the reciprocal lattice is
$\mathbf{G}_1=(\frac{2\pi}{a},0)$, $\mathbf{G}_2=(0,\frac{2\pi}{a})$.
The band structure of Eq. \ref{eq:Ham0} is calculated by using the
plane-wave basis.

In the absence of SO coupling, the two-component bosons with strong optical
potentials can be described by the lattice Bose-Hubbard model as
\bea
H_{Hub}=-\sum_{\langle ij\rangle,\sigma}
t_{ij} \big[b^\dag_{i,\sigma}b_{j,\sigma}+h.c \big]
+\sum_{i}[\frac{U}2 n_i^2-\mu n_i],
\eea
where $\sigma=\uparrow,\downarrow$ denote the pseudospin components;
$b_{i\sigma}$ and $b^\dag_{i\sigma}$ are bosonic annihilation and creation
operators for spin $\sigma$ at site $i$, respectively.
$\sum_{\langle i,j\rangle}$ denotes the summation over all the
nearest neighbors. $n_i$ is the boson density operator at site $i$:
$n_i=\sum_\sigma b^\dag_{i\sigma}b_{i\sigma}$.
Generally, the interaction can be spin-dependent.
In this article, we only consider the spin-independent interaction.

We first consider the case of $\alpha=1$, $\beta=0$, which
is the situation directly related with current experiments in
the absence of optical lattice \cite{lin2011}.
The SO coupling induces an extra term in the tight-binding term as
\bea
H_{so}=-\lambda\sum_\mathbf{i}[b_{\mathbf{i},\uparrow}^\dag
b_{\mathbf{i}+\vec{e}_x,\downarrow}-b_{\mathbf{i},\downarrow}^\dag
b_{\mathbf{i}+\vec{e}_x,\uparrow}]+h.c,
\label{eq:SO1}
\eea
where $\vec{e}_x$ is the unit vector along the $x$-direction.
In momentum space, Eq. \ref{eq:SO1} becomes
$H_{so}=\sum_\mathbf{k}
\Psi_\mathbf{k}^\dag \hat{H}_\mathbf{k}\Psi_\mathbf{k}$, where
$\Psi_\mathbf{k}=[b_{\mathbf{k},\uparrow},b_{\mathbf{k},\downarrow}]^T$,
and $\hat{H}^1_\mathbf{k}$ is a 2 by 2 matrix reads as:
\bea
\hat{H}_\mathbf{k}=\varepsilon_\mathbf{k} \hat{I}+2\lambda \sin k_x
\hat{\sigma}_y
\label{eq:momentum}
\eea
where $\varepsilon_\mathbf{k}=-2(t_x \cos k_x+ t_y \cos k_y)-\mu$
and $t_{x} (t_y)$ is the hopping integrals along $x$ and $y$-directions,
respectively.
In the long-wave limit $k\rightarrow 0$, Eq. (\ref{eq:momentum}) reduces to
the Hamiltonian in continuous space realized in experiments.

The SO coupling is equivalent to a pure gauge at $\beta=0$, which can be
eliminated by a gauge transformation
\bea
 U=\exp\{i\mathbf{k_{so}}\cdot
\mathbf{r} \sigma_z\},
\label{eq:gauge}
\eea
which applies to the doublet $(b_\uparrow,b_\downarrow)$.
The energy spectra of Eq. \ref{eq:momentum} has two branches
as $E_{\pm}= -2t_y[\cos (k_x\pm k_{so})
+\cos k_y]-\mu$ corresponding to the eigenvalues $\pm 1$ of $\sigma_y$,
and the following relations are satisfied
\bea
t_x=t_y\cos k_{so}, \ \ \  \lambda=t_y \sin k_{so}.
\eea
Bosons condense into the energy minima of
$\pm \mathbf{Q}=(\pm k_0,0)$ with $k_1=\arctan(\lambda/t_y)$.
The corresponding single particle wavefunctions at these two minima are:
\bea
\Psi_{\pm\mathbf{Q}}=\frac{1}{\sqrt{2}}e^{\pm i\mathbf{r}\cdot\mathbf{Q}_{sc}}
\left(\begin{array}{c}1
\\\pm i\end{array}
\right).
\eea
At the Hartree-Fock level, bosons can take either of $\Psi_{\mathbf{Q}_{sc}}$
as a plane-wave spin-polarized state, or, a superposition of
them as $\frac{1}{\sqrt 2} (\Psi_{\mathbf{Q}} + \Psi_{\mathbf{Q}})=
[\cos \mathbf Q \cdot \mathbf{r}, \sin \mathbf Q \cdot \mathbf{r}]^T$
with the same energy.
The latter one can be stabilized by spin-dependent interaction
of $H_{sp,int}=U^\prime \sum (n_{i,\uparrow} -n_{i,\downarrow})^2$
with $U^\prime<0$.
It exhibits a spin spiral states in the $xz$-plane with the pitch
wavevector $2\mathbf{Q}$ as plotted in Fig. \ref{fig:1D}.
We will see that in the Mott-insulating state, although strong
interaction suppresses the superfluidity, the spin configuration
remains the same spiral order.

\begin{figure}[htb]
\includegraphics[width=0.95\linewidth]{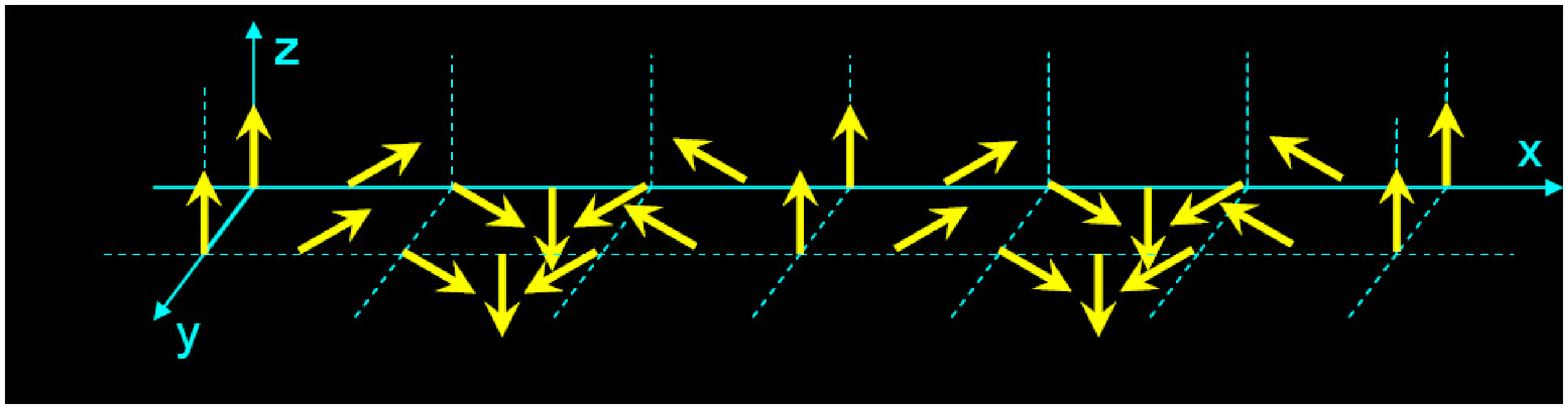}
\caption{Spin spiral configurations of the Bose-Hubbard model
with unidirectional SO coupling.
It is valid for both the incommensurate superfluid state,
and the Mott insulating state.}
\label{fig:1D}
\end{figure}

We consider the Mott insulating state at $\langle n_i\rangle=1$, and
construct the superexchange Hamiltonian for the pseudospin-$\frac{1}{2}$
bosons as
\bea
H_{eff}=\sum_\mathbf{i} [H_{\mathbf{i},\mathbf{i}+\hat{e}_y}+
H_{\mathbf{i},\mathbf{i}+\hat{e}_x}].
\label{eq:eff1}
\eea
For the vertical bond without SO coupling, $H_{i,i+e_y}$ is just the $SU(2)$
ferromagnetic Heisenberg superexchange \cite{duan2003,kuklov2003} as
$H_{\mathbf{i},\mathbf{i}+\hat{e}_y}=-J_{1,y} \mathbf{S}_{\mathbf{i}}
\cdot\mathbf{S}_{\mathbf{i}+\hat{e}_y}$ where $J_{1,y}=4t_y^2/U>0$.
For the horizontal bond,  the SO coupling
leads to the Dzyaloshinsky-Moriya (DM) type superexchange
terms \cite{dzyaloshinsky1958,moriya1960} as
\bea
H_{\mathbf{i},\mathbf{i}+\hat{e}_x}&=&-J_{1,x}
\mathbf{S}_{\mathbf{i}}\cdot\mathbf{S}_{\mathbf{i}+\hat{e}_x}-
J_{12}\mathbf{d}_{\mathbf{i},\mathbf{i}+\hat{e}_x}\cdot
(\mathbf{S}_{\mathbf{i}}\times\mathbf{S}_{\mathbf{i}+\hat{e}_x})
\nn \\
&+&J_2[\mathbf{S}_{\mathbf{i}}\cdot\mathbf{S}_{\mathbf{i}+\hat{e}_x}
-2(\mathbf{S}_{\mathbf{i}}\cdot\mathbf{d}_{\mathbf{i},\mathbf{i}+\hat{e}_x})
(\mathbf{S}_{\mathbf{i}+\hat{e}_x}\cdot\mathbf{d}_{\mathbf{i},
\mathbf{i}+\hat{e}_x})],\nn \\
\label{eq:DM1}
\eea
where $J_2=4\lambda^2/U$, $J_{12}=4t_y\lambda/U$.
$\mathbf{d}_{\mathbf{i},\mathbf{i}+\hat{e}_x}$ is a 3D DM vector defined
on the bond $[\mathbf{i},\mathbf{i}+\hat{e}_x]$, and,
$\mathbf{d}_{\mathbf{i},\mathbf{i}+\hat{e}_x}=\hat{e}_y$.

The DM term of Eq.(\ref{eq:DM1}) prefers a spin spiral ordering along
the horizontal direction, as illustrated in Fig.\ref{fig:1D} (b).
The effect of the gauge transformation Eq. \ref{eq:gauge}
on spin operators is to rotate
$\mathbf{S}_{\mathbf{i}}$ around $y$-axis at the angle
of $2m\theta$ where $m$ is the horizontal coordinate of site $i$
and $\theta=\arctan(\lambda/t_y)$ \cite{shekhtman1992}, such that
\bea
\mathbf{S}'_{\mathbf{i}}&=&(1-\cos2m\theta)[\mathbf{d}\cdot
\mathbf{S}_\mathbf{i}]\mathbf{d}+\cos2m\theta\mathbf{S}_\mathbf{i}\nn \\
&-&\sin2m\theta ~\mathbf{S}_\mathbf{i}\times\mathbf{d},
\label{eq:rotation}
\eea
where $\mathbf{d}=\mathbf{d}_{\mathbf{i},\mathbf{i}+\hat{e}_x}=\hat{e}_y$.
Through this transformation, the DM interaction is gauged away, and
Eq.(\ref{eq:DM1}) turns into a ferromagnetic coupling:
\bea
H_{\mathbf{i},\mathbf{i}+\hat{e}_x}=-J_0\mathbf{S}'_{\mathbf{i}}\cdot\mathbf{S}'_{\mathbf{i}+\hat{e}_x},
\label{eq:ferro2}
\eea
where $J_0=J_{1,y}=4(t_x^2+\lambda^2)/U$.
The the exchange model becomes an isotropic ferromagnetic Heisenberg
model, and thus spin polarization can point along any direction.
In order to obtain the actual spin spiral configuration, we need
to do the inverse operation of  Eq. \ref{eq:rotation}.
Say, if we choose the classic spin at the point of origin
along $z$ direction $\mathbf{S}_{[0,0]}=\hat{e}_z$, according to the
rotation defined in Eq.(\ref{eq:rotation}), all the spins in the
classic ground state are restricted within the $x$-$z$ plane, and
the classic spin at the point $[m,n]$ is
$\mathbf{S}_{[m,n]}=\cos(2m\theta)\hat{e}_z+\sin(2m\theta)\hat{e}_x$.
As shown in Fig.\ref{fig:1D} (b), the classic spins form a chiral
pattern with a characteristic length, which is the same
as in the superfluid case as plotted in Fig. \ref{fig:1D}.
The only difference is that the superfluid phase coherence is
lost in the Mott-insulating state.

\begin{figure}[htb]
\includegraphics[width=0.7\linewidth]{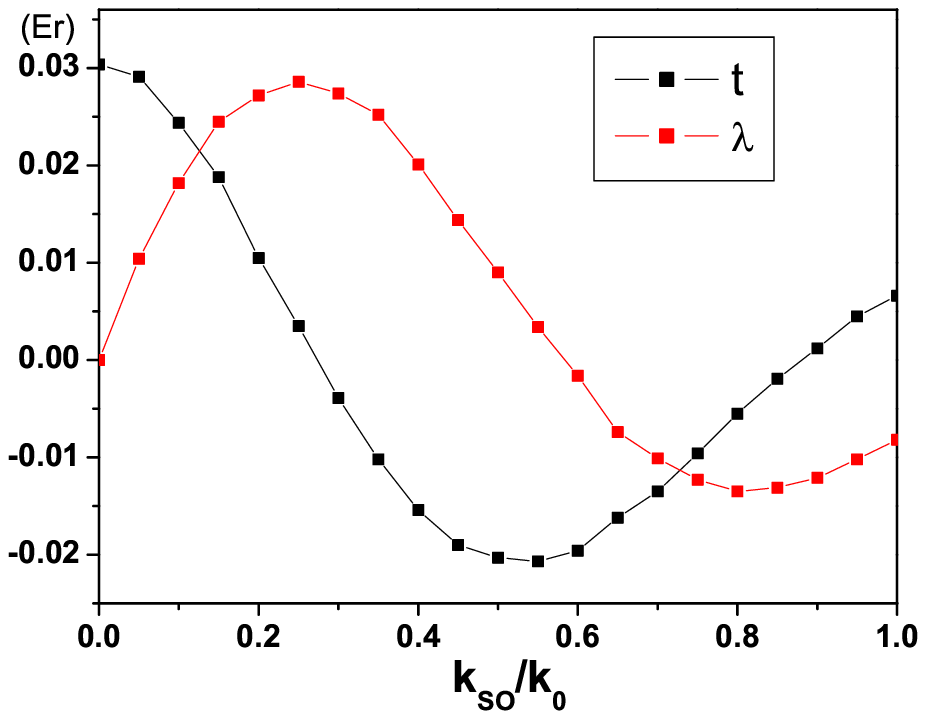}
\includegraphics[width=0.70\linewidth]{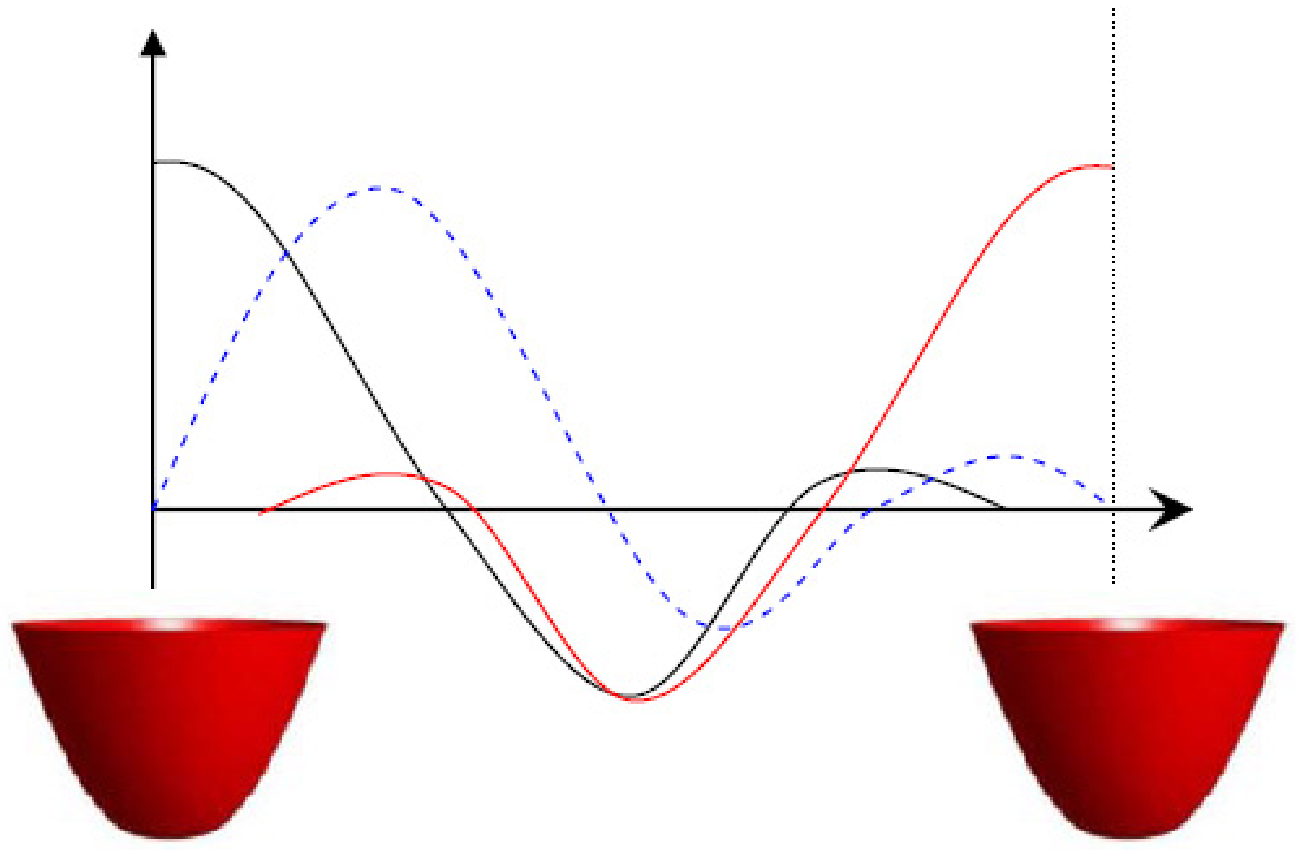}
\caption{A) The dependence of the spin-independent hopping integral
$t$ and the spin-dependent one $\lambda$ {\it v.s.} the SO coupling
strength $\gamma=k_{so}/k_0$.
The optical potential depth is $V_0=8E_r$.
B) Sketch of Wannier wavefunctions for $f(r)$ (solid black line)
and $g(r)$ (dashed blue line) in Eq. \ref{eq:wannier}.
}
\label{fig:parameter}
\end{figure}

Now we discuss the isotropic Rashba SO coupling with $\alpha=\beta=1$.
From the symmetry analysis, we easily have $t_x=t_y$ for spin
independent hoppings, while the spin-dependent SO hoppings become
\bea
H'_{SO}&=&-\lambda\sum_\mathbf{i}[b_{\mathbf{i},\uparrow}^\dag
b_{\mathbf{i}+\vec{e}_x,\downarrow}-b_{\mathbf{i},\downarrow}^\dag
b_{\mathbf{i}+\vec{e}_x,\uparrow}]+h.c, \nn \\
&-&i\lambda\sum_\mathbf{i}[b_{\mathbf{i},\downarrow}^\dag
b_{\mathbf{i}+\vec{e}_y,\uparrow}+b_{\mathbf{i},\uparrow}^\dag
b_{\mathbf{i}+\vec{e}_y,\downarrow}]+h.c  .
\label{eq:SO2}
\eea
In momentum space, the tight-binding band Hamiltonian turns to:
$H'=\sum_\mathbf{k} \Psi_\mathbf{k}^\dag
\hat{H}'_\mathbf{k}\Psi_\mathbf{k}$, where
\bea
\hat{H}'_\mathbf{k}=\varepsilon_\mathbf{k} \hat{I}+2\lambda [\sin
k_x \hat{\sigma}_y+\sin k_y \hat{\sigma}_x],
\label{eq:momentum2}
\eea
where $\varepsilon_\mathbf{k}=-2t (\cos k_x x + \cos k_y y) $.
The energy spectra of Eq.(\ref{eq:momentum2}) read
\begin{equation}
E'_{\pm}=\varepsilon_\mathbf{k}\pm2\lambda\sqrt{\sin^2k_x+\sin^2k_y}.
\label{eq:Energytb}
\end{equation}
The band minima are four-fold degenerate at the points $\mathbf{Q}_{sc}=(\pm
k, \pm k)$, where $k=\arctan\frac{\lambda}{\sqrt{2}t}$.

Next we calculate the band parameters $t$ and $\lambda$ versus SO
coupling parameter $\gamma$, by fitting the band spectra using the
plane-wave basis in the continuum.
The results are plotted in  Fig. \ref{fig:parameter} A.
Both $t$ and $\lambda$ oscillate and decay as increasing $\gamma$,
which can be understood from the behavior of the onsite Wannier functions.
Each optical site can be viewed as a local harmonic potential
and the lowest single particle state wavefunction was calculated in
Ref. [\onlinecite{wu2008}]
\bea
\psi_{j_z=\frac{1}{2}}(\vec r)= [f(r), g(r) e^{i \phi}]^T,
\label{eq:wannier}
\eea
and its time-reversal partner is
$\psi_{j_z=-\frac{1}{2}}(\vec r)=(-g(r) e^{i \phi}, f(r))$.
$f(r)$ and $g(r)$ are real radial wavefunctions, which exhibit characteristic
oscillations with the pitch value $k_{so}$ and a relative phase shift
approximately $\frac{\pi}{2}$ as plotted in Fig. \ref{fig:parameter} B.
$t$ and $\lambda$ are related to the off-centered integrals
of $f(r)$ and $g(r)$ of two sites, which overlap in the middle.
As a result, $t$ and $\lambda$ also oscillate as increasing
$\gamma$, which also exhibit a phase shift approximately
at $\pi/2$ as shown in Fig. \ref{fig:parameter} A.

We would like to clarify one important and subtle point.
Actually the on-site Wannier functions are no-longer spin eigenstates,
but total angular momentum eigenstate $j_z=\frac{1}{2}$, and thus
are still a pair of Kramer doublets.
For the operators $(b_{i\uparrow}, b_{i,\downarrow})^T$ defined on
site $i$,  they do not refer to
spin eigenbasis but to the $j_z$-eigenbasis.
In fact, in the case that $k_{so}\ge k_0$, the onsite spin moments
are nearly zero.
The $j_z$-movements mainly come from orbital angular momentum.
As pointed out in Ref. [\onlinecite{wu2008}], the Wannier functions
of $j_z$ eigenstates exhibit skyrmion-type spin texture distributions
and half-quantum vortex on each site.
This phenomena also remind us of the Friedel oscillation in
solid state physics.
In the case of $k_{so}\gg k_0$, each site exhibits
Landau level-type quantization:
states with different values of $j_z$ are nearly degenerate
\cite{wu2008,hu2012}, and a single band picture ceases to work here.

\begin{figure}[htb]
\includegraphics[width=0.43\linewidth]{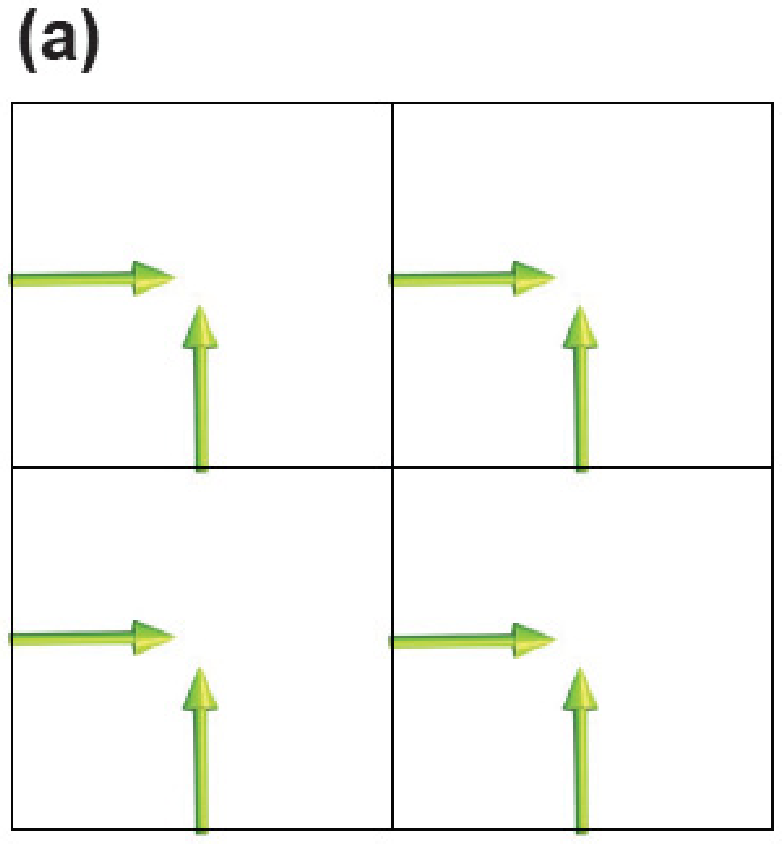}
\includegraphics[width=0.51\linewidth]{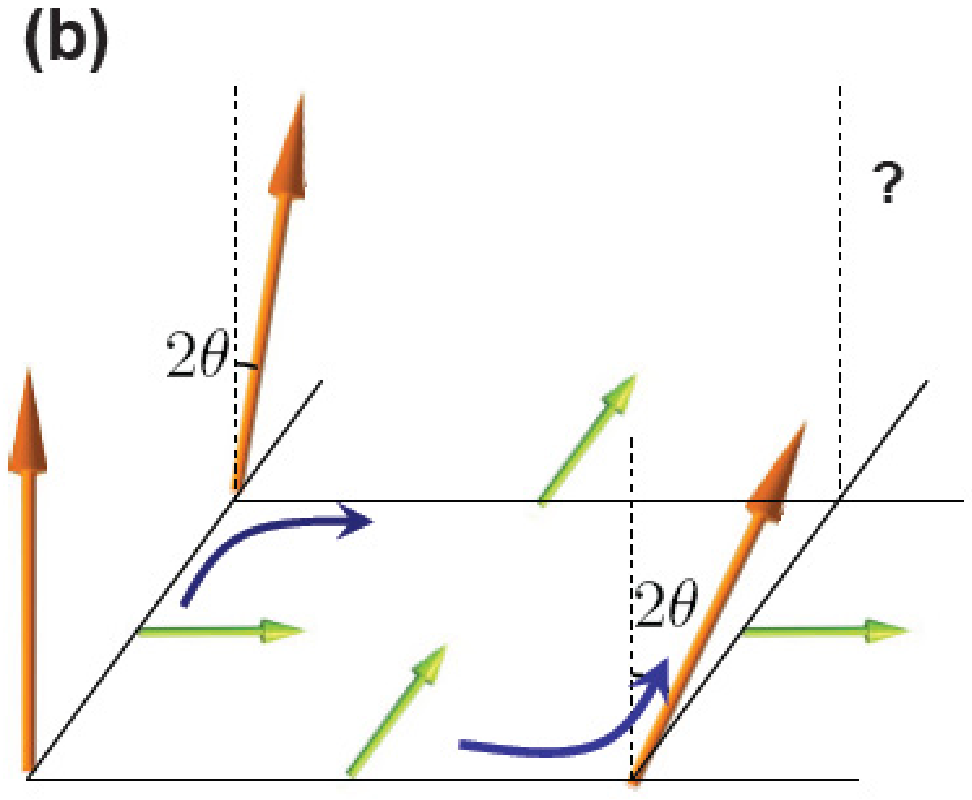}
\caption{(A) The pattern of DM vectors of the superexchange magnetic
model in the Mott-insulating state. (B) Illustration of the
frustration in the spin configuration with  DM interactions, the
rotations around $x$-axis and $y$-axis do not commutate with each
other.} \label{fig:DM}
\end{figure}

Deep inside the Mott-insulating phase, we obtain the effective magnetic
Hamiltonian:
\bea
H'_{eff}=\sum_\mathbf{i} [H'_{\mathbf{i},\mathbf{i}+\hat{e}_y}+
H'_{\mathbf{i},\mathbf{i}+\hat{e}_x}].
\label{eq:eff2}
\eea
$H'_{\mathbf{i},\mathbf{i}+\hat{e}_x}$ is the same as Eq.(\ref{eq:DM1}), and
\bea
\nonumber H'_{\mathbf{i},\mathbf{i}+\hat{e}_y}=-J_1
\mathbf{S}_{\mathbf{i}}\cdot\mathbf{S}_{\mathbf{i}+\hat{e}_y}
-J_{12}\mathbf{d}_{\mathbf{i},\mathbf{i}+\hat{e}_y}\cdot
(\mathbf{S}_{\mathbf{i}}\times\mathbf{S}_{\mathbf{i}+\hat{e}_y})
\\+J_2[\mathbf{S}_{\mathbf{i}}\cdot\mathbf{S}_{\mathbf{i}+\hat{e}_y}
-2(\mathbf{S}_{\mathbf{i}}\cdot\mathbf{d}_{\mathbf{i},\mathbf{i}+\hat{e}_y})
(\mathbf{S}_{\mathbf{i}+\hat{e}_y}\cdot\mathbf{d}_{\mathbf{i},\mathbf{i}+\hat{e}_y})],
\label{eq:DM2}
\eea
where $\mathbf{d}_{\mathbf{i},\mathbf{i}+\hat{e}_y}=\hat{e}_x$.
The pattern of the DM vectors is shown in Fig.\ref{fig:DM} (a), which is
a strongly reminiscent of that in cuprate superconductor
YBa$_2$Cu$_3$O$_6$ \cite{coffey1991,bonesteel1993}.
The classical ground state of Eq. (\ref{eq:eff2}) is nontrivial because
the DM interaction can not be gauged away:
DM vectors in horizontal bonds favor spiraling around the $y$-axis, while
that in vertical bonds favor spiraling around the $x$-axis.
Since rotations around $x$ and $y$-axis do not commune,
no spin configurations can simultaneously satisfy both requirements,
which leads to spin frustrations shown in Fig. \ref{fig:DM} (B).

The quantum situation of Eq. \ref{eq:eff2} is even more involved,
which can only be solved approximately.
Below we focus on two situations: $|\lambda|\ll |t|$ and $|\lambda|\gg |t|$.
At $\lambda=0$, the ground state of Eq. (\ref{eq:eff2}) is known to be
ferromagnetism.
At $|\lambda/t|\ll 1$, we use spin-wave approximation to
analyze the instability of a ferromagnetic state induced by the
DM interaction.
Notice that in this case, it is impossible to find a global rotation as
in Eq. (\ref{eq:rotation}) to gauge away the DM
vectors and transform Eq. (\ref{eq:eff2}) to an $SO(3)$ invariant
Hamiltonian, thus the quantized axis in the spin wave analysis can
not be chosen arbitrarily. To gain some insight, we choose a classic
ferromagnetic state as a variational ground state
parameterized by
$\mathbf{S}_{\mathbf{i}}=S(\cos\gamma\sin\eta,\sin\gamma\sin\eta,\cos\eta)$.
The corresponding variational energy  $E_0=-S^2(J_1-J_2+2\sin^2\eta
J_2)$ is minimized when $\eta=\pi/2$, which implies that
the $xy$-plane is the easy plane.

\begin{figure}[htb]
\includegraphics[width=0.6\linewidth] {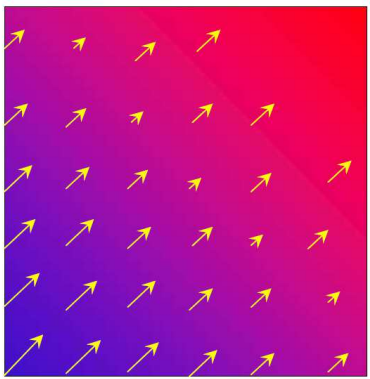}
\caption{(Spin spiral ordering in the limit of $|\lambda|\ll |t|$.}
\label{fig:spiral}
\end{figure}

To calculate the spin waves spectra, it is convenient to rotate the
coordinate so that the new $z$-axis points along the direction
$\mathbf{l}=[\bar{1}\bar{1}0]$ in the original coordinate (we choose
$\mathbf{l}$ as the quantized axis).
The Holstein-Primakoff transformation is employed to transform
Eq. (\ref{eq:eff2}) into the bosonic Hamiltonian:
\bea
H_{b}=\sum_{i,\mu}-J_0(\cos2\theta-i\sin2\theta/\sqrt{2})a_i^\dag
a_{i+e_\mu}+h.c,
\eea
where $\theta=\arctan(\lambda/t)$ as defined above, $\mu=x,y$.
We only keep quadric terms and ignore the terms proportional to
$\sin^2\theta$ since $\lambda/t\ll 1$.
In momentum space, it becomes
\bea
H'_{ex}&=&-2J_0\sum_{\mathbf{k}}[\cos2\theta\cos
k_x+\frac{1}{\sqrt{2}}\sin2\theta\sin k_x \nn \\
&+&\cos2\theta\cos k_y+\frac{1}{\sqrt{2}}\sin2\theta\sin k_y]
~ c^\dag_\mathbf{k} c_\mathbf{k},
\label{eq:spectrum2}
\eea
The minimum of the dispersion of Eq. \ref{eq:spectrum2} occurs at
points $\mathbf{Q}_M=(\pm k', \pm k')$, where $k'$ satisfies that $\tan
k'=\frac{1}{\sqrt{2}}\tan 2\theta$.
Compare it with the energy minima in the noninteracting band
Hamiltonian $\mathbf{Q}_{sf}=(\pm k,\pm k)$, we have $k^\prime=2k$
at the limit of $\gamma\rightarrow 0$.
The nonzero minimum of the magnon spectrum is a signature of the
spin spiral order, as shown in Fig. \ref{fig:spiral} (a).

Interestingly, in the opposite limit of $|\lambda/t|\gg 1$, Eq.
(\ref{eq:DM1}) can be related to that of $|\lambda/t|\ll 1$ through
a duality transformation. On site $i$ with the coordinates $(i_x,
i_y)$, $\vec S_i$ is transformed into \bea
&&S_{i_x,i_y}^x\rightarrow (-1)^{i_x}\mathbb{S}_{i_x,i_y}^x; \ \ \,
S_{i_x,i_y}^y\rightarrow
(-1)^{i_y}\mathbb{S}_{i_x,i_y}^y;\nn \\
&&S_{i_x,i_y}^z\rightarrow (-1)^{i_x+i_y}\mathbb{S}_{i_x,i_y}^z.
\label{eq:duality} \eea $\mathbb{\vec S}_i$ still maintains the spin
commutation relation. Under this transformation, the $J_1$-term
transforms into the $J_2$-term and vice versa, and the $J_{12}$-term
is invariant. Thus this dual transformation indicates that there is
a one-to-one correspondence between the $J_2$-dominant phase
($|\lambda/t|\gg 1$) and that of $J_1$ with $|\lambda/t|\ll 1$ which
has been analyzed above.

In conclusion, we have investigated the magnetic ordering of
two-component Bose-Hubbard model with synthetic SO coupling. The
band parameters of hopping integrals exhibit characteristic
oscillations as increasing SO coupling strength, and the onsite
magnetic moments are nearly orbital moments at large SO coupling
strength. In the Mott-insulating state with one particle per site,
an effective magnetic superexchange model with the DM type
interaction is derive. The spin spiral state and its dual state are
found in the limits of $|\lambda|\ll |t|$ and $|\lambda|\gg |t|$.

This work was supported by the NSF DMR-1105945, the AFOSR-YIP program.

{\it Note added}
Up the posting of this paper, we become aware two papers on the
similar topic \cite{radic2012,cole2012}.


\begin{thebibliography}{99}

\bibitem{read1991}
N. Read and S. Sachdev, Phys. Rev. Lett. {\bf 66}, 1773 (1991).

\bibitem{sachdev1991}
S. Sachdev and N. Read, Mod. Phys. Lett. {\bf 5}, 219 (1991).

\bibitem{fulde1964}
P. Fulde and R. A. Ferrell, Phys. Rev. {\textbf 135}, A550 (1964).

\bibitem{larkin1965} A. I. Larkin and Y. N. Ovchinnikov, Sov. Phys.-JETP {\bf 20}, 762
(1965).

\bibitem{feynman1972}
R. P. Feynman, {\it Statistical Mechanics, A Set of Lectures}
(Berlin: Addison-Wesley, 1972).

\bibitem{leggett2001}
A. J. Leggett, Rev. Mod. Phys. 73, 307 (2001).

\bibitem{wu2008}
C. Wu , I. Mondragon-Shem, arXiv:0809.3532; C. Wu , I.
Mondragon-Shem, and X. F. Zhou, Chin. Phys. Lett., {\bf 28}, 097102
(2011).

\bibitem{high2012}
A.A. High {\it et al.}, Nature 483, 584 (2012). A.A. High {\it et
al.}, arXiv:1103.0321.

\bibitem{lin2009} Y.-J. Lin et al., Nature \textbf{462}, 628 (2009).

\bibitem{lin2011} Y.-J. Lin, K. Jim\'{e}nez-Garc\'{\i}a and I. B. Spielman,
Nature \textbf{471}, 83 (2011).

\bibitem{stanescu2008} T. Stanescu, B. Anderson, V. Galitski
Phys. Rev. A 78, 023616 (2008).

\bibitem{ho2011} T.-L. Ho and S. Zhang, Phys. Rev. Lett.
\textbf{107}, 150403 (2011).

\bibitem{wang2010}  C. Wang, C. Gao, C.M. Jian, H. Zhai,
Phys. Rev. Lett. \textbf{105}, 160403 (2010).

\bibitem{yip2011}  S.-K. Yip, Phys. Rev. A \textbf{83}, 043616 (2011).

\bibitem{zhang2012} Y. Zhang, L. Mao, and C. Zhang, Phys. Rev. Lett.
\textbf{108}, 035302 (2012).

\bibitem{zhou2011} X.-F. Zhou, J. Zhou, and C. Wu, Phys. Rev. A \textbf{84},
063624 (2011).

\bibitem{li2012a} Y. Li, X. F. Zhou, and C. Wu, arXiv:1205.2162.

\bibitem{hu2012} H. Hu, B. Ramachandhran, H. Pu, and X.J Liu,
Phys. Rev. Lett. \textbf{108}, 010402(2012).

\bibitem{mondragon2010}
I. Mondragon-shem, B. A. Rodriguez, C. Wu, Bull. Am. Phys. Soc.
{\bf 55},  MAR.Z31.11 (2010).
I. Mondragon-shem, Bachelor thesis,
Instituto de FA­sica, Facultad de Ciencias Exactas y Naturales,
Universidad de Antioquia (2010).


\bibitem{santos2011}  S. Sinha, R. Nath, and L. Santos, arXiv:1102.2045.


\bibitem{li2012} Y. Li, X. Zhou, and C. Wu, Phys. Rev. B \textbf{85},
125122 (2012).

\bibitem{ghosh2011} S.K. Ghosh, J.P. Vyasanakere, V.B. Shenoy,
Phys. Rev. A, \textbf{84}, 053629 (2011).





\bibitem{dzyaloshinsky1958}
I. Dzyaloshinsky, J. Phys. and Chem. Sol. {\bf 4}, 241 (1958).

\bibitem{moriya1960}
T. Moriya, Phys. Rev. {\bf 120}, 91 (1960).

\bibitem{duan2003}
L. M. Duan, E. Demler, and M. D. Lukin, Phys. Rev. Lett.
{\bf 91}, 090402 (2003).

\bibitem{kuklov2003}
A. B. Kuklov and B. V. Svistunov, Phys. Rev. Lett.
{\bf 90}, 100401 (2003).

\bibitem{shekhtman1992}
L. Shekhtman, O. Entin-Wohlman, and A. Aharony, Phys.
Rev. Lett. 69, 836 (1992).

\bibitem{coffey1991}
D. Coffey, T. M. Rice, and F. C. Zhang, Phys. Rev. B 44,
10112 (1991).

\bibitem{bonesteel1993}
N. E. Bonesteel, Phys. Rev. B 47, 11302 (1993).

\bibitem{radic2012}
J. Radic, A. Di Ciolo, K. Sun, V. Galitski,
arXiv:1205.2110.

\bibitem{cole2012}
W. S. Cole, S. Z. Zhang, A. Paramekanti, and N Trivedi,
arXiv:1205.2319.


\end{thebibliography}
\end{document}